\documentclass{amsart}
\usepackage{times}
\usepackage{amssymb,amsmath,amscd, amsthm,stmaryrd,verbatim, mathrsfs}

\usepackage[colorlinks,linkcolor=blue,urlcolor=cyan,citecolor=green,pagebackref]{hyperref}

\usepackage{color}

\newtheorem{Thm}{Theorem}
\newtheorem{theorem}{Theorem}[section]

\newtheorem{Lem}[theorem]{Lemma}

\newtheorem{Prop}{Proposition}[section]

\newtheorem{Definition}{Definition}

\numberwithin{equation}{section}

\newcommand{\mb}{\mathbf}

\newcommand{\mr}{\mathrm}
\newcommand{\frk}{\mathfrak}

\usepackage{hyperref}
\begin{document}

\title{Particle Motion in Generalized Dirac's Monopoles of dimension $2k+1$ }

\author{Zhanqiang Bai}

\email{zhanqiang.bai@whu.edu.cn}

\address{School of Mathematics and Statistics, Wuhan University, Wuhan 430072, China }

\maketitle
\begin{abstract}
By using Meng's idea in his generalization of the classical MICZ-Kepler problem, we obtained the equations of motion of a  charged particle in the field of generalized  Dirac monopole in odd
dimensional Euclidean spaces.
The main result is that for every particle trajectory $\mathbf{r}:I\to\mathbb{R}^{2k+1}\setminus\{0\}$, there is
a 2-dimensional cone with vertex at the origin on which $\mathbf{r}$ is a~geodesic.


 \end{abstract}


\section {Introduction}
In 1896, Henri Poincar\'e \cite{Po} investigated the following one-body dynamical system
\begin{eqnarray} \mathbf{r}''=\lambda \frac{\mathbf{r}\times\dot{\mathbf{r}}}{|\mathbf{r}|^3},\end{eqnarray}
for $\mathbf{r}\in\dot{\mathbb{R}}^3:=\mathbb{R}^3\setminus\{0\}$ and a~constant $\lambda\in\mathbb{R}$. He used this model to give an explanation  for an experiment of the physicist Kristian Birkeland, which consisted of approaching one
pole of a strong magnet near cathode rays, the other pole being far enough to be considered
negligible. In 1930, Dirac \cite{Di} used this model to study  the problem of the classical motion of an electrically charged particle in the field of Dirac's magnetic monopole. Since then, this system has been  thoroughly studied by many people, see Ref. \cite{Feher,Fierz,Goddard,Haas,Horvathy,Jackiw,Lapidus,Moreira,Ritter,Sivardiere}. Poincar\'e firstly showed that for every solution $\mathbf{r}$, there is a~cone with vertex
at the origin on which $\mathbf{r}$ is a~geodesic.
Moreover, he provided an explicit expression for the cone's direction and the angle at its vertex (which vary
depending on the initial conditions and the charge~$\lambda$).

Recently, Mayrand \cite{Mayrand} showed that particles in Yang's monopole (which is a natural generalization of Dirac's monople in $\mathbb{R}^5$) follow geodesics on 2-
dimensional cones, and hence all solutions can be obtained explicitly.

In 2007, Meng\cite{Meng2007-1} found that Dirac's monopole can be generalized in  all Euclidean spaces $\mathbb{R}^n$. Since then, Meng\cite{Meng2007-2,Meng2008-1, Meng2009-1,Meng2010-1, Meng2011-1,Meng2013-1} began to reconstruct and generalize the MICZ-Kepler problem (which is the magnetized version of the classical Kepler problem). In 2013, we \cite{BaiMengWang} found that for all odd dimensions, the orbits of the generalized MICZ-Kepler problems are all conics.  By using the same method with  \cite{BaiMengWang}, in this paper we find that the particles in all odd dimensional Dirac's monopoles will follow geodesics on cones.

\subsection{Outline}

In section 2, we derive the equations of motion of the Poincar\'{e} problem in odd dimensions. In section 3, we show that a charged particle in odd dimensional generalized Dirac's monopole must follow a geodesic on a cone.

 \subsection{Notations and Conventions}
We are mainly dealing with poly-vectors in the Euclidean space $\mathbb R^n$.

Let $V$ be  $\mathbb R^n$ and  $k>0$ be an integer. A $k$-vector in $V$ is just an element of $\wedge^k V$. A $k$-vector is called {\em decomposable} if it is the wedge product of $k$ vectors. It is a trivial fact that a $2$-vector in a 3D space is always decomposable. In case $X$ is a non-zero decomposable $k$-vector in $V$, we use $[X]$ to denote the $k$-dimensional oriented subspace of $V$, with $X$ representing its orientation.

The inner product extend from vectors to poly-vectors and is denoted  by $(, )$. By definition, for vectors $u_1$, \ldots, $u_k$, $v_1$, \ldots, $v_k$ in $V$, let $[u_i\cdot v_j]$ be the square matrix whose $(i,j)$-entry is $u_i\cdot u_j$, then
$$
( u_1\wedge\cdots\wedge u_k, v_1\wedge\cdots\wedge v_k) =\det [u_i\cdot v_j].
$$

We define the interior product $\lrcorner$ as the adjoint of the wedge product with respect to the inner product for poly-vectors: for poly-vectors $X$, $u$ and $v$ in $V$ with $\deg X+\deg u=\deg v$, we have
$$
( X\wedge u, v\rangle= \langle u, X\lrcorner v).
$$

For poly-vector $X$, we write $X^2$ for $( X, X )$. In case $( X, X )\ge 0$, we write $|X|$ for $\sqrt{\langle X, X\rangle}$. We always write $r$ for $|\mathbf r|$. 

\section{The generalized Poincar\'{e} problem in dimension $n=2k+1$}\label{S: Kepler}
The purpose of this section is to describe the equation of motion for the  generalized Poincar\'{e} problem in odd dimension $n=2k+1\ge5$. This equation is the $n$-dimensional analogue of the equation of motion
\begin{eqnarray}\label{EM}
{\mathbf{r}}=\lambda\frac{\mathbf{r}\times\dot{\mathbf{r}}}{|\mathbf{r}|^3}
\end{eqnarray} for the Poincar\'{e} problem in $\mathbb R^3$, where the parameter $\lambda$ is the magnetic charge.

Firstly we recall Meng's construction for the generalized Dirac monopoles in high dimensions. The details can be found in Meng \cite{Meng2007-1,Meng2013}.

We let $G=\mathrm{SO}(2k)$ and consider the canonical principal $G$-bundle over $\mathrm{S}^{2k}$:
$$\begin{array}{c}
 \mathrm{SO}(2k+1)\cr
 \Big\downarrow\cr
 \mathrm{S}^{2k}.
\end{array}
 $$This bundle comes with a natural connection
$$\omega (g):=\mathrm{Pr}_{\frk{so}(2k)}\left(g^{-1}dg\right),$$
where $g^{-1}dg$ is the Maurer-Cartan form for $ \mathrm{SO}(2k+1)$, so it is an $\frk{so}(2k+1)$-valued differential one form on $\mathrm{SO}(2k+1)$, and $\mathrm{Pr}_{\frk{so}(2k)}$ denotes the orthogonal projection of $\frk{so}(2k+1)$ onto $\frk{g}:=\frk{so}(2k)$.

Under the map
\begin{eqnarray}
\pi:  \mathbb R^{2k+1}_* & \to & \mathrm{S}^{2k}\cr
 \mb  r &\mapsto &{\mb r \over r},
\end{eqnarray}
the above bundle and connection are pulled back to a principal $G$-bundle
\begin{eqnarray}
\begin{array}{l}
P\cr
 \Big\downarrow\cr
X:=\mathbb R^{2k+1}_*
\end{array}
\end{eqnarray} with a connection which is usually referred to as the generalized Dirac monopole \cite{Meng2007-1}. Now $$P_\lambda\to \mathbb R^{2k+1}_*$$ is the associated fiber bundle with fiber being a certain co-adjoint orbit $\mathcal O_\lambda$ of $G$, the so-called magnetic orbit  with magnetic charge $\lambda\in \mathbb R$.

To describe $\mathcal O_\lambda$, let us use $\gamma_{ab}$ ($1\le a, b\le 2k$) to denote the element of $i\frk{g}$ such that in the defining representation of $\frk{g}$, $M_{a,b}:=i\gamma_{ab}$ is represented by the skew-symmetric real symmetric matrix whose $a b$-entry is $-1$, $b a$ entry is $1$, and all other entries are $0$. For the invariant metric $(, )$ on $\frk{g}$, we take the one such that $M_{a,b}$ ($1\le a< b\le 2k$) form an orthonormal basis for $\frk g$. Via this invariant metric, one can identify $\frk{g}^*$ with $\frk{g}$, hence co-adjoint orbits with adjoint orbits. By definition, for any $\lambda\in \mathbb R$,
\begin{eqnarray}
\mathcal O_\lambda := \mr{SO}(2k)\cdot {1\over \sqrt k}(|\lambda|M_{1,2}+\cdots + |\lambda| M_{2k-3, 2k-2} + \lambda M_{2k-1, 2k} ).
\end{eqnarray} It is easy to see that $\mathcal O_\lambda=\{0\}$ if $\lambda=0$ and is diffeomorphic to $\mr{SO}(2k)\over \mr{U}(k)$ if $\lambda\neq 0$.

\vskip 5pt

We shall let the small Lartin letters $j$, $k$, etc. be indices that run from $1$ to $n$, and the small Latin letters $a$, $b$, etc. be indices that run from $1$ to $n-1$. To do local computations, we need to choose a bundle trivialization on $U$ which is $\mathbb R^n$ with the negative $n$-th axis removed and then write down the gauge potential explicitly. Indeed, there is a bundle trivialization on $U$ such that the gauge potential $A=A_k\, dx^k$ can be written as
\begin{eqnarray}\label{mnple}
A_n =0,\quad A_b=-{1\over r(r+x_n)}x^a\gamma_{ab}
\end{eqnarray}
where $x^a\gamma_{ab}$ means $\sum_{a=1}^{n-1}x^a\gamma_{ab}$, something we shall assume whenever there is a repeated index.  It is then clear that the gauge field strength
$F_{jk}: = \partial_jA_k-\partial_k A_j + i [A_j, A_k]$ is of the form
\begin{eqnarray}\label{Ffield}
F_{nb} = {1\over r^3}x^a\gamma_{ab},\quad
F_{ab} = -{1\over r^2}(\gamma_{ab} +x^aA_b - x^bA_a).
\end{eqnarray}

The following lemma follows from Meng\cite{Meng2013}.
\begin{Lem}\label{lemma}
Assume $n=2k+1$. Let $Q={1\over 2k}\sum_{a, b}(\gamma_{ab})^2$ and $\nabla_k=\partial_k+iA_k$. For the gauge potential $A$ defined in Eq. (\ref{mnple}),  the following statements are true.

1) As $i\,\frk{so}(2k)$-valued functions on $U$,
\begin{eqnarray}\label{Id1}
x^k A_k=0,\quad x^j F_{jk}=0,\quad \nabla_l F_{jk}={1\over r^2}\left( -x^j
F_{lk}-x^k F_{jl}-2x^l F_{jk} \right).
\end{eqnarray}
Consequently, we have $\mathbf r\lrcorner F=0$ and ${D\over dt} (r^2F)=-\mathbf r\wedge (\mathbf r'\lrcorner F)$.

2) Assume $\xi\in \mathcal O_\lambda\subset \frk{g}$. As real functions on $U$,
\begin{eqnarray}\label{Id3}
r^4\sum_k( i\xi, F_{kj})(i \xi, F_{kj'})={\lambda^2\over k}\left(\delta_{jj'}-{x^j x^{j'}\over
r^2}\right).
\end{eqnarray}  Consequently, we have $|(i\xi, r^2F)|^2=\lambda^2$ and $$(i\xi, \mathbf r' \lrcorner F)\lrcorner (i\xi, r^2F)=-{\lambda^2/k\over r}\left({\mathbf r\over r}\right)', \quad |r^2(i\xi, \mathbf r' \lrcorner F)|^2={\lambda^2\over k}{|\mathbf r\wedge \mathbf r'|^2\over r^2}.$$
\end{Lem}

We are now ready to describe the equation of motion for the generalized Poincar\'{e} problem in dimension  $2k+1$. Let $\mathbf r$: $\mathbb R\to X$ be a smooth map, and $\xi$ be a smooth lifting of $\mathbf r$:
\begin{eqnarray}
\begin{array}{ccc}
 & & P_\lambda\\
 \\
 & \xi \nearrow &\Big\downarrow\\
 \\
 \mathbb R &\buildrel\mathbf r\over\longrightarrow & X
\end{array}
\end{eqnarray}
Let $Ad_P$ be the adjoint bundle $P\times_G\frk{g}\to X$, $d_\nabla$ be the canonical connection, i.e., the generalized Dirac monopole on $\mathbb R^{2k+1}_*=X$. Then the curvature $\Omega:=d_\nabla^2$ is a smooth section of the vector bundle $\wedge^2T^*X\otimes Ad_P$. (With a trivialization of $P\to X$, locally $\Omega$ can be represented by ${1\over 2}\sqrt{-1}F_{jk}\, dx^j\wedge dx^k$.)
The equation of motion is
\begin{eqnarray}\label{EqnMF}
\fbox{$\left\{
\begin{array}{l}
\mathbf r'' = (\xi,\mathbf r'\lrcorner \Omega),\cr
\\
{D\over dt}\xi =0.
\end{array}
\right.$}
\end{eqnarray} Here ${D\over dt}\xi$ is the covariant derivative of $\xi$, $(, )$ refers to the inner product on the fiber of the adjoint bundle coming from the invariant inner product on $\frak g$, and 2-forms are identified with 2-vectors via the standard Euclidean structure of $\mathbb R^{2k+1}$. Eq. (\ref{EqnMF}) defines a super integrable model, referred to as the {\bf generalized Poincar\'{e} problem  with magnetic charge $\lambda$ in dimension $2k+1$}, which generalize the Poincar\'{e} problem with magnetic charge $\lambda$ in dimension 3. Indeed, in dimension $3$, the bundle is topological trivial, $\xi=\lambda M_{12}$, and $\Omega = {*(\sum_{i=1}^3 x^i\, dx^i)\over r^3} M_{12}$, then Eq. (\ref{EqnMF}) reduces to Eq. (\ref{EM}), i.e., the equation of motion for the Poincar\'{e} problem with magnetic charge $\lambda$ in dimension 3.

\section{The angular momentum}\label{S: Lenz}

Let $\mathbf{r} \in \mathbb{R}^{2k+1}_{\ast}$, and we denote
\begin{eqnarray}
L=\mathbf r\wedge \mathbf r'+ (\xi,r^2\Omega),
\end{eqnarray}

called the angular momentum  of the generalized Poincar\'{e} problem.

\begin{Lem}
$L$ is a constant of motion.
\end{Lem}
\begin{proof}
To verify that $L$ is a constant of motion directly, it suffices to do local computations over $U$, where $U$ is $\mathbb R^{2k+1}$ with a so called Dirac string (i.e., the negative $(2k+1)$-th coordinate axis) removed, and, for a given non-colliding orbit, by dimension reason we may assume it misses the entire $(2k+1)$-the coordinate axis.

Over the dense open set $U$, the bundle $P\to X$ shall be trivialized in a way so that the gauge potential is of the form as in equation (\ref{mnple}), so that the curvature $\Omega$ shall be represented by $iF$ and the lifting $\xi$ shall be represented by a smooth map (also denoted by $\xi$) from $\mathbb R$ into $\frk g$ whose image is always inside $\mathcal O_\lambda$.  With this understood,  we have
\begin{eqnarray}
L' &= & \mathbf r\wedge \mathbf r''+(i\xi, -\mathbf r\wedge (\mathbf r' \lrcorner F))\cr
&= & \mathbf r\wedge (\mathbf r''-(\xi, \mathbf r' \lrcorner \Omega))\cr
&=&  \mathbf r\wedge 0=0.\nonumber
\end{eqnarray}

\end{proof}

For a non-colliding orbit, since $\mathbf r\wedge \mathbf r'\neq 0$, use the lemma, we have
\begin{eqnarray}\label{Lmu}
|L|^2= |\mathbf r\wedge \mathbf r'|^2+ |(i\xi, r^2F)|^2=  |\mathbf r\wedge \mathbf r'|^2+\lambda^2> \lambda^2.
\end{eqnarray}

A solution $(\mathbf r (t), \xi(t))$ to the equation of motion (\ref{EqnMF}) shall be referred to as a {\em motion}, whose total trace inside $P_\lambda$ shall be referred to as an {\em orbit}. Under the bundle projection $\pi_{\lambda}: P_\lambda\to\mathbb R^{2k+1}_*$, these orbits become curves inside the ``external configuration space" $\mathbb R^{2k+1}_*$, with the non-colliding ones being referred to as {\em generalized Poincar\'{e} orbits}. In this section we shall show that the generalized Poincar\'{e} orbits are all geodesics on conics.

Let $(\mathbf r (t), \xi(t))$ be a motion that represents a generalized Poincar\'{e} orbit, and
\begin{eqnarray}V:= \mathbf r\wedge\mathbf r'\wedge r^3\mathbf r''=\mathbf r\wedge\mathbf r'\wedge r^3(i\xi, \mathbf r'\lrcorner F) .
\end{eqnarray} Since both $\mathbf r$ and $\mathbf r'$ are orthogonal to $r^3(i\xi, \mathbf r'\lrcorner F)$, we have
\begin{eqnarray}\label{Vnormsq}
|V|^2=|\mathbf r\wedge\mathbf r'|^2\; |r^3(i\xi, \mathbf r'\lrcorner F)|^2={\lambda^2\over k}|\mathbf r\wedge\mathbf r'|^4={\lambda^2\over k}(|L|^2-\lambda^2)^2
\end{eqnarray} by Lemma \ref{lemma} and Eq. (\ref{Lmu}). Therefore $|V|$ is a constant of motion. In view of Eq. (\ref{Lmu}), it vanishes if and only if $\lambda=0$.

\begin{Lem} The $3$-vector $V$ in $\mathbb R^{2k+1}$ is a constant of motion, and it vanishes if and only if $\lambda=0$.
\end{Lem}
\begin{proof} Using the product rule for differentiation, we have
\begin{eqnarray}
V'&=&   \mathbf r\wedge\mathbf r'\wedge ((r  \mathbf r')'\lrcorner (i\xi, r^2F))+ \mathbf r\wedge\mathbf r'\wedge (r \mathbf r' \lrcorner (i\xi, r^2F)')\cr
&=& \mathbf r\wedge\mathbf r'\wedge ((r  \mathbf r')'\lrcorner (i\xi, r^2F))+ \mathbf r\wedge\mathbf r'\wedge (r\mathbf r' \lrcorner (i\xi, {D\over dt}(r^2F)))\quad \mbox{by eqn of motion}\cr
&=& \mathbf r\wedge\mathbf r'\wedge ((r  \mathbf r')'\lrcorner (i\xi, r^2F))+ \mathbf r\wedge\mathbf r'\wedge  (r\mathbf r' \lrcorner (i\xi, -\mathbf r\wedge (\mathbf r'\lrcorner F)))\quad \mbox{by Lemma \ref{lemma}}\cr
&=& \mathbf r\wedge\mathbf r'\wedge ((r  \mathbf r')'\lrcorner (i\xi, r^2F))-\mathbf r\wedge\mathbf r'\wedge r^2r' (i\xi, \mathbf r'\lrcorner F) \cr
&=& \mathbf r\wedge\mathbf r'\wedge (r  \mathbf r''\lrcorner (i\xi, r^2F))\cr
&=& \mathbf r\wedge\mathbf r'\wedge (r (i\xi, \mathbf r'\lrcorner F))\lrcorner (i\xi, r^2F))\quad \mbox{by eqn of motion and Lemma \ref{lemma}}\cr
&=& -{\lambda^2\over k}\mathbf r\wedge\mathbf r'\wedge \left({\mathbf r\over r}\right)'\quad \mbox{by Lemma \ref{lemma}}\cr
&=& 0.\nonumber
\end{eqnarray}
The rest is clear.
\end{proof}

The nonzero constant decomposable $3$-vector $V$ determines a constant subspace $[V]$ of $\mathbb R^{2k+1}$. Since $\mathbf r\wedge V=0$, we know that the generalized Poincar\'{e} orbit is inside the 3D space $[V]$, in fact a geodesic on a cone inside $[V]$, as we shall see in a moment.

Let $\bar L$ be the image of $L$ under the orthogonal projection $\wedge^2 {\mathbb R}^{2k+1}\to \wedge^2 [V]$. Being referred to as the {\bf effective angular momentum}, $\bar L$ shall be seen to play an important role in the study of generalized Poincar\'{e} problems.
A simple computation shows that
\begin{eqnarray}\label{Lbar}
{\bar L}=\left(\mathbf{r}-\frac{r^4}{|L|^2-\mu^2} (i\xi, \mathbf{r}'\lrcorner F)\right)\wedge \left(\mathbf{r}'-\frac{r'}{r} \mathbf{r}\right),
\end{eqnarray}  a decomposable $2$-vector inside the 3D space $[V]$.

From the definition, $\bar L$ should be a constant of motion, a fact which can be verified directly. This fact holds even when $\lambda=0$ because $\bar L=L$ in this special case.  By some simple computations, one can verify that $\mathbf r\in [V]$.

\section{The  main theorem}
First of all, we review some definitions and propositions from Mayrand\cite{Mayrand}.

\begin{Definition}
\label{DefConeBasic}
The \emph{cone} of aperture $\psi\in(0,\pi/2]$ directed along $\mathbf{L}\in\dot{\mathbb{R}}^{n+1}$ is the set of all
points $\mathbf{r}\in\dot{\mathbb{R}}^{n+1}$ satisfying
\begin{gather}
\label{ConeDef}
\frac{\mathbf{r}\cdot\mathbf{L}}{|\mathbf{r}||\mathbf{L}|}=\cos\psi.
\end{gather}
We write an ``$n$-dimensional cone in $\mathbb{R}^{n+1}$''  for any such
cone.
\end{Definition}

\begin{Definition}
\label{DefConeGen}
Let~$P$ be an af\/f\/ine~$m$-dimensional plane in $\mathbb{R}^{n+1}$ that intersects with $S^{n}$ in more than one point.
The \emph{cone generated by~$P$ in $\mathbb{R}^{n+1}$} is the set of all points $\mathbf{r}\in\dot{\mathbb{R}}^{n+1}$ such
that $\mathbf{r}/|\mathbf{r}|\in P$.
We write a~``$m$-dimensional cone in $\mathbb{R}^{n+1}$'' for any such set.
\end{Definition}

\begin{Prop}
\label{TheoremConeGeo}
Let $\mathbf{r}:I\to\dot{\mathbb{R}}^{n+1}$ be a~non-colliding curve, where $n\geq 2$.
Then, $\mathbf{r}$ is a~geodesic on an~$n$-dimensional cone~$C$ if and only if $\mathbf{r}$ is a~geodesic on
a~$2$-dimensional cone $D\subseteq C$ of the same aperture.
\end{Prop}

\begin{Prop}\label{cone} Let $C$ be the cone of aperture $\psi$ directed along $L \in\dot{\mathbb{R}}^{n+1}$ and let $V$ be a $m$-dimensional subspace containing $L$. Then, $C\cap V$ is a $(m-1)$-dimensional cone of aperture $\psi$.

\end{Prop}

The following is our main theorem which extends Theorem 7.1 of Ref. \cite{Mayrand}.
\begin{Thm} \label{maintheorem1}

Let $(\mathbf r , \xi)$ be a solution to the equation of motion (\ref{EqnMF}) in $\mathbb{R}^{2k+1}$. If $\mathbf{r}$ is non-colliding (i.e., $\mathbf r\wedge \mathbf r'\neq 0$), then $\mathbf{r}$ is  a geodesic on the cone of aperture $\psi=arccos(\frac{|\lambda|}{\sqrt{k}|\bar L|})$ directed along $\star\bar{ L}$, where $\star$ is the Hodge star operator in the 3D space $[V]$.
\end{Thm}
\begin{proof}
From the definition of $\bar L$, we know it is a constant two form. And $[V]$ is a constant 3D vector space. Also we have
\begin{align*}|\bar L|^2&=|\mathbf{r}-\frac{r^4}{|L|^2-\mu^2} (i\xi, \mathbf{r}'\lrcorner F)|^2  \cdot | \mathbf{r}'-\frac{r'}{r} \mathbf{r}|^2-(\mathbf{r}-\frac{r^4}{|L|^2-\mu^2} (i\xi, \mathbf{r}'\lrcorner F),\mathbf{r}'-\frac{r'}{r} \mathbf{r})^2\\
&=|\mathbf{r}\wedge \mathbf{r}'|^2+\frac{\lambda^2}{k}=|L|^2-|\lambda|^2+\frac{\lambda^2}{k}>\frac{\lambda^2}{k}.
\end{align*}

So we can define a Hodge star operator
$$\star: \wedge^{2}[V]\rightarrow [V],$$
which is a linear isomorphism between the two vector spaces and $\frac{V}{|V|}$ is the preferred unit 3-vector. The inverse of $\star$ is also a linear isomorphism from $[V]$ to  $\wedge^{2}[V]$. We still denote it by $\star$.  Then $\star (\bar L)$ is a constant vector in $[V]$ and $\star\star (\bar L)=\bar L$.

From the equation \ref{Lbar}, we have
\begin{equation}\frac{\mathbf{r}}{r}\wedge \frac{\bar L}{|\bar L|}=\frac{V}{|\bar L|(|L|^2-\lambda^2)}=\frac{|\lambda|}{\sqrt{k}|\bar L|}\frac{V}{|V|}. \end{equation}

From the definition of Hodge star operator, we have
\begin{equation}\frac{\mathbf{r}}{r}\wedge \star\star (\frac{\bar{ L}}{|\bar L|})=(\frac{\mathbf{r}}{r}, \star (\frac{\bar{ L}}{|\bar L|})\frac{V}{|V|}.\end{equation}

 So from theses two equations, we must have
 $$(\frac{\mathbf{r}}{r}, \star (\frac{\bar{ L}}{|\bar L|})=\frac{|\lambda|}{\sqrt{k}|\bar L|}\in (0,1),$$
which is  also a constant.
Then we can find some angle $\psi\in (0, \pi/2)$, such that $cos\psi= \frac{|\lambda|}{\sqrt{k}|\bar L|}.$
So $\mathbf{r}$ lies on the cone $C$ of aperture $\psi=arccos(\frac{|\lambda|}{\sqrt{k}|\bar L|})$ directed along $\star\bar{ L}$.  From proposition \ref{cone}, we know  $\mathbf{r}$ lies on the $2$-dimensional cone $C\cap [V]$.

Also we have $$(\mathbf{r}'', \mathbf{r})=((i\xi, \mathbf r'\lrcorner F),\mathbf{r})=((i\xi,  F), \mathbf r' \wedge  \mathbf{r})=-((i\xi, \mathbf r\lrcorner F),\mathbf{r}')=0,$$
and $$(\mathbf{r}'', \mathbf{r}')=((i\xi, \mathbf r'\lrcorner F),\mathbf{r}')=((i\xi,  F), \mathbf r' \wedge  \mathbf{r}')=0.$$
Thus $\mathbf{r}''$ is orthogonal to $\mathbf{r}$ and $\mathbf{r}'$. Then form corollary 6.1 of Ref. \cite{Mayrand}, we know $\mathbf{r}$ is a geodesic on the cone of aperture $\psi=arccos(\frac{|\lambda|}{\sqrt{k}|\bar L|})$ directed along $\star\bar{ L}$.

\end{proof}

\subsection*{Acknowledgements}

The author is grateful to Professor Guowu Meng for introducing this problem to the author.
The author would also like to thank Maxence Mayrand for helpful discussions.

\end{document}